\documentclass[referee]{raa}   

\usepackage{graphicx,times}
\usepackage{natbib}
\usepackage{amssymb,amsmath}

\bibpunct{(}{)}{;}{a}{}{,}

\begin{document}

\title{Four New Eclipsing Binary Systems in the TESS field: CD-34~13220, HD~295082, TYC~6484-426-1 and TYC~6527-2310-1
}

 \volnopage{ {\bf 20XX} Vol.\ {\bf X} No. {\bf XX}, 000--000}
   \setcounter{page}{1}

   \author{Burak Ula\c{s}\inst{1,2}
   }

   \institute{ Department of Space Sciences and Technologies, Faculty of Arts and Sciences, \c{C}anakkale Onsekiz Mart University, Terzio\v{g}lu Campus, TR~17100, \c{C}anakkale, Turkey; {\it burak.ulas@comu.edu.tr}\\
        \and
             Astrophysics Research Centre and Observatory, \c{C}anakkale Onsekiz Mart University, Terzio\v{g}lu Campus, TR~17100, \c{C}anakkale, Turkey\\
\vs \no
   {\small Received 20XX Month Day; accepted 20XX Month Day}
}

\abstract{We present the first evidence for the binarity of four targets in the TESS field. The temperatures are estimated by SED analysis and the orbital periods are determined. The TESS light curves of the systems are analysed and the orbital and the absolute parameters are derived. The targets are also compared to well-studied binary systems with the same morphological type and their evolutionary states are discussed. Our results indicate that the stars belong to the class of eclipsing detached binary systems. 
\keywords{binaries: eclipsing --- stars: fundamental parameters --- stars: individual: CD-34~13220, HD~295082, TYC~6484-426-1, TYC~6527-2310-1 
}
}

   \authorrunning{B. Ula\c{s}}            
   \titlerunning{Four New Eclipsing Binaries}  
   \maketitle

%
\section{Introduction}           
\label{sect:intro}

Binary stars are significant stellar objects in the sense of allowing the researchers to determine the physical parameters of the components precisely and investigate the stellar evolution of the stars in various circumstances. Ever since Giovanni B. Riccioli's observation of Mizar in the 17th century and Christian Mayer's first catalog of double stars in 1781, astronomers forge away in understanding the light variation and physical properties of these types of systems. The number of newly discovered binary star systems were increased at the beginning of the 1900s, the years when the astronomers were keen on finding new systems, as mentioned in \cite{nie01}. At present time, a recently updated version of Binary Star Database (\citealt{kov15}) consists of about 120000 stellar systems. A growing number of space-based observations also made a major contribution to the discovery and investigation of the properties of the component stars in detail, thanks to their precise data. Eclipsing binaries, a subclass of binary star systems, are also important tools in astrophysics in understanding the structure and characteristics of the stars. The analyses of the light and radial velocity curves of eclipsing binaries enable us to derive not only the physical parameters like mass, radius, luminosity, and temperature but also structural phenomena just as apsidal motion and atmospheric parameters such as limb darkening, gravity darkening as remarked by \cite{gui93}. 

Our information on several stellar phenomena, including binarity, was increased by recent space-based missions like {\it Kepler} \cite{bor10} and Transiting Exoplanet Survey Satellite (TESS, \citealt{ric15}). For instance, Kepler Eclipsing Binary Catalog (\citealt{kir16}) consists of light curves and the related parameters of 2878 systems. TESS, on the other hand, is very opportune in studying low amplitude variations such as the light curves of pulsating stars in binary systems. The mission focused on the identification of transiting planets around nearby stars by scanning the 85\% of the sky in its first two years (\citealt{ric15}). The several parameters of approximately 400000 selected stars are planned to be cataloged in TESS Input Catalog (TIC, \citealt{sta19}). In addition to the above programs, ground-based surveys like The Large Sky Area Multi-Object Fiber Spectroscopic Telescope (LAMOST, \citealt{zha12}) and The All Sky Automated Survey (ASAS, \citealt{poj02}) provided a great improvement in the discovery and the knowledge of the structure and the evolution of binary stars. For example, the atmospheric parameters of 2020 EA- and 9149 EW-type binary systems observed by LAMOST were cataloged, and their evolutionary states were discussed by \cite{qia18} and \cite{qia20}. 256000 binary or variable star candidates were also detected by \cite{qia19} among 786400 stars falling into the field of LAMOST.

The targets investigated in this study are not classified as eclipsing binary systems in the literature so far. CD-34~13220 (TIC~277373390) can be found in several catalogs (\citealt{tho94, nat93, mor01, cut03, ros94, hog00, gai18})
which list the position and magnitude information of the stars substantially. \cite{cru19} list HD~295082 (TIC~42828781) in their Mid-infrared Stellar Diameters and Fluxes Compilation Catalogue with the estimated distance and angular diameter values of 533.2~pc and 0.063~mas, respectively. The star also appears in The Henry Draper Extension Charts (\citealt{nes95}) in which the spectral class, position, and magnitude values are tabulated. TYC~6484-426-1 (TIC~31303242) is in the photometric empirical calibration research of \cite{rui18}. They included the star to the list of targets with low extinction by giving $E(B-V)_{max}$ = 0.0083. \cite{can18} remarked that TYC~6527-2310-1 (TIC~63192395) is not belonging to an open cluster in their study on astrometric parameters of 128 open clusters. The star was also listed among the solar-like oscillators in the TESS field by \cite{sch19} which gives fundamental parameters combining the Gaia DR2 and Hipparcos data.  

The lack of comprehensive studies in the literature motivated us to investigate the systems in detail. In the next section, we give the properties of the data and features of the TESS light curves. The details of the light curve analyses and the results proving our targets' binarity are explained in Sec~\ref{sec:lcan}. The last section deals with the discussion of the results, evolutionary status, and comparison of the systems to well-known binary stars from previous studies.

\section{Light Curve Data}
\label{sect:lcdat}

\begin{figure}[htbp]
\centering
\includegraphics[scale=0.75]{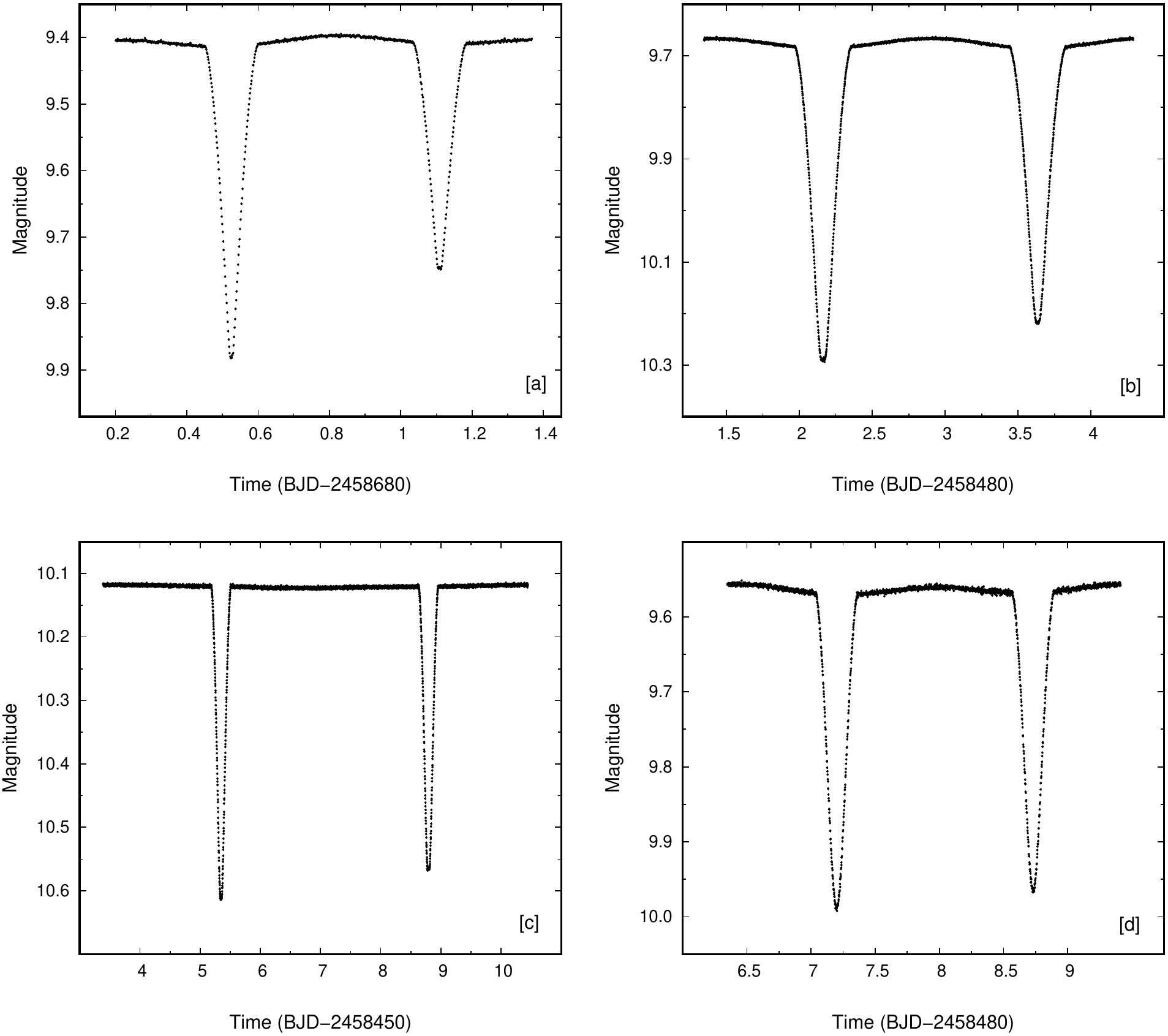}
\caption{One-period-long light curves of CD-34~13220 (a), HD~295082 (b), TYC~ 6484-426-1 (c), and TYC~6527-2310-1 (d). Magnitudes are calculated by using the TESS magnitudes given in \cite{sta19}.}
\label{figoplc}
\end{figure}

The TESS light curves of the systems were acquired from the light curve files that were achieved through {\sc MAST} (Mikulski Archive of Space Telescope Portal)\footnote{mast.stsci.edu/portal/Mashup/Clients/Mast/Portal.html}. The files contain data that the photometric analysis and cotrending were applied (\citealt{tan18}). The PDC\_SAP (Pre-search Data Conditioning Simple Aperture Photometry) flux values ($F_i$) were extracted from the data files and converted to magnitudes ($m_i$) by using the equation $m_{i} = -2.5 \log F_{i}$. The TESS magnitudes (9.45$^m$, 9.74$^m$, 10.14$^m$, and 9.60$^m$ for CD-34~13220, HD~295082, TYC~6484-426-1, and TYC~6527-2310-1, respectively) given by \cite{sta19} were used during the derivation of the magnitudes. A linear trend as a function of time was also removed from the data of TYC~6484-426-1 and TYC~6527-2310-1. Fig.~\ref{figoplc} shows the systems' light variations in one orbital period intervals. These curves represent the common eclipsing binary type variations.

The depth of the primary and secondary minimums are 0.49$^m$ and 0.35$^m$ for CD-34~13220, 0.63$^m$ and 0.55$^m$ for HD~295082, 0.50$^m$ and 0.45$^m$ for TYC~6484-426-1 and 0.44$^m$ and 0.41$^m$ for TYC~6527-2310-1. Primary minimum lasts 3.8$^h$, 9.8$^h$, 9.6$^h$, and 8.4$^h$ for CD-34~13220, HD~295082, TYC~6484-426-1, and TYC~6527-2310-1, respectively, while the durations of the secondary minimums are 3.6$^h$, 9.1$^h$, 9.6$^h$, and 8.2$^h$. The data are from observation sector 13 for CD-34~13220, sector 6 for HD~295082, sector 5 and 6 for TYC~6484-426-1, and sector 6 and 7 for TYC~6527-2310-1. Furthermore, the lack of minima times in the literature directed us to calculate times of minimum lights for the systems using the method of \cite{kwe56}. Table~\ref{tabmin} lists 92 calculated times of primary and secondary minimums. In the table, TYC~ 6484 and TYC~6527 refer to TYC~ 6484-426-1 and TYC~6527-2310-1, respectively.

The shape and the characteristics of the light curves verge upon the light variation of a common detached eclipsing binary system. Therefore, detached morphology must be considered during the analyses, as we introduce it in the next section.

\begin{table}[htbp]
\caption{\label{tabmin}Calculated times of minimum light for the systems} 
\small
 \begin{tabular}{lccclcclcc}
  \hline\noalign{\smallskip}
System & BJD-2457000 & Type&System & BJD-2457000 & Type&System & BJD-2457000 & Type\\
\hline
CD-34~13220~~&1657.169086(5)&I&&1677.021844(5)&I&&1476.51675(1)&I\\
&1656.585189(7)&II&&1677.605866(7)&II&&1483.57357(1)&I\\
&1658.336961(5)&I&&1678.189751(5)&I&&1479.96552(1)&II\\
&1657.753109(9)&II&&1678.774260(9)&II&&1487.02209(1)&II\\
&1659.504652(5)&I&&1679.357437(5)&I&TYC~6527&1468.81026(1)&I\\
&1658.920870(7)&II&&1679.941601(7)&II&&1470.34182(1)&II\\
&1660.671987(5)&I&&1680.525091(5)&I&&1471.87524(1)&I\\
&1660.088789(7)&II&&1681.109487(7)&II&&1473.40676(1)&II\\
&1661.840223(5)&I&HD~295082~&1470.39149(1)&I&&1474.94005(1)&I\\
&1661.256646(7)&II&&1468.91985(1)&II&&1476.47153(1)&II\\
&1663.008070(5)&I&&1473.33431(1)&I&&1481.06986(1)&I\\
&1662.424379(7)&II&&1471.86274(1)&II&&1479.53642(1)&II\\
&1664.175831(6)&I&&1476.27736(1)&I&&1484.13472(1)&I\\
&1663.592091(7)&II&&1474.80601(1)&II&&1482.60138(1)&II\\
&1664.760032(7)&II&&1479.22043(1)&I&&1487.19953(1)&I\\
&1666.511393(5)&I&&1480.69188(1)&II&&1485.66590(1)&II\\
&1665.927797(7)&II&&1482.16348(1)&I&&1493.32929(1)&I\\
&1668.847170(5)&I&&1483.63491(1)&II&&1491.79568(1)&II\\
&1669.431362(7)&II&&1485.10649(1)&I&&1496.39400(1)&I\\
&1670.014987(5)&I&&1486.57796(1)&II&&1494.86020(1)&II\\
&1670.599053(7)&II&&1488.04936(1)&I&&1499.45889(1)&I\\
&1671.182855(5)&I&&1489.52086(1)&II&&1497.92537(2)&II\\
&1671.766854(7)&II&TYC~6484&1441.23240(2)&I&&1502.52388(1)&I\\
&1672.350583(5)&I&&1444.68152(1)&II&&1500.99050(1)&II\\
&1672.934813(7)&II&&1448.28973(1)&I&&1505.58870(1)&I\\
&1673.518278(5)&I&&1455.34646(1)&I&&1507.11994(1)&II\\
&1674.102572(7)&II&&1451.73825(1)&II&&1508.65350(1)&I\\
&1674.686146(5)&I&&1462.40318(1)&I&&1510.18469(1)&II\\
&1675.270310(7)&II&&1458.79505(1)&II&&1511.71839(1)&I\\
&1675.854005(5)&I&&1469.45999(1)&I&&1513.24959(1)&II\\
&1676.438160(7)&II&&1472.90868(1)&II& & & \\
\hline
\end{tabular}
\end{table}

\section{Analyses of the Light Curves}
\label{sec:lcan}

One of the most critical parameters to achieve realistic results in the light curve analysis is the effective temperature of the primary component. Since there is very limited information on the systems in the literature, we estimated the temperatures of the primaries by constructing spectral energy distributions (SED) through Virtual Observatory SED Analyzer (VOSA, \citealt{bay08}) for the available photometric data in the VizieR database (\citealt{och00}). Parameter-grid search was used to fit the photometric data to achieve the optimized Kurucz atmosphere model (\citealt{kur79}). The process was done for each system in the present study and the results were assumed to be the effective temperature of the primary components.

The absence of sufficient knowledge of the orbital properties of systems in the literature led us to determine orbital period values to be used in the light curve analyses. For doing this, we first employed Fourier analysis to the light curve data of each system using {\tt PERIOD04} software (\citealt{len05}) and obtained a frequency value that corresponds to the orbital period. Alternatively, we averaged the differences between consecutive time of minimum lights of the same types to estimate a period value. From those two values, which constructs the better light curve was adopted as the orbital period for each system. Period analyses were also conducted to times of minimum in Table~\ref{tabmin} by assuming a linear trend. Variation in time of minimum light ($\Delta T_0$) and orbital period ($\Delta P$) were calculated using the equation $O-C=\Delta T_{0}+\Delta P\cdot E$, where $E$ is the cycle number. Results are listed for each system in the last two lines of Table~\ref{tablc} and Fig.~\ref{figlc1} and \ref{figlc2} are plotted by using the final period values. 

\begin{figure}[htbp]
   \centering
\includegraphics[scale=0.8]{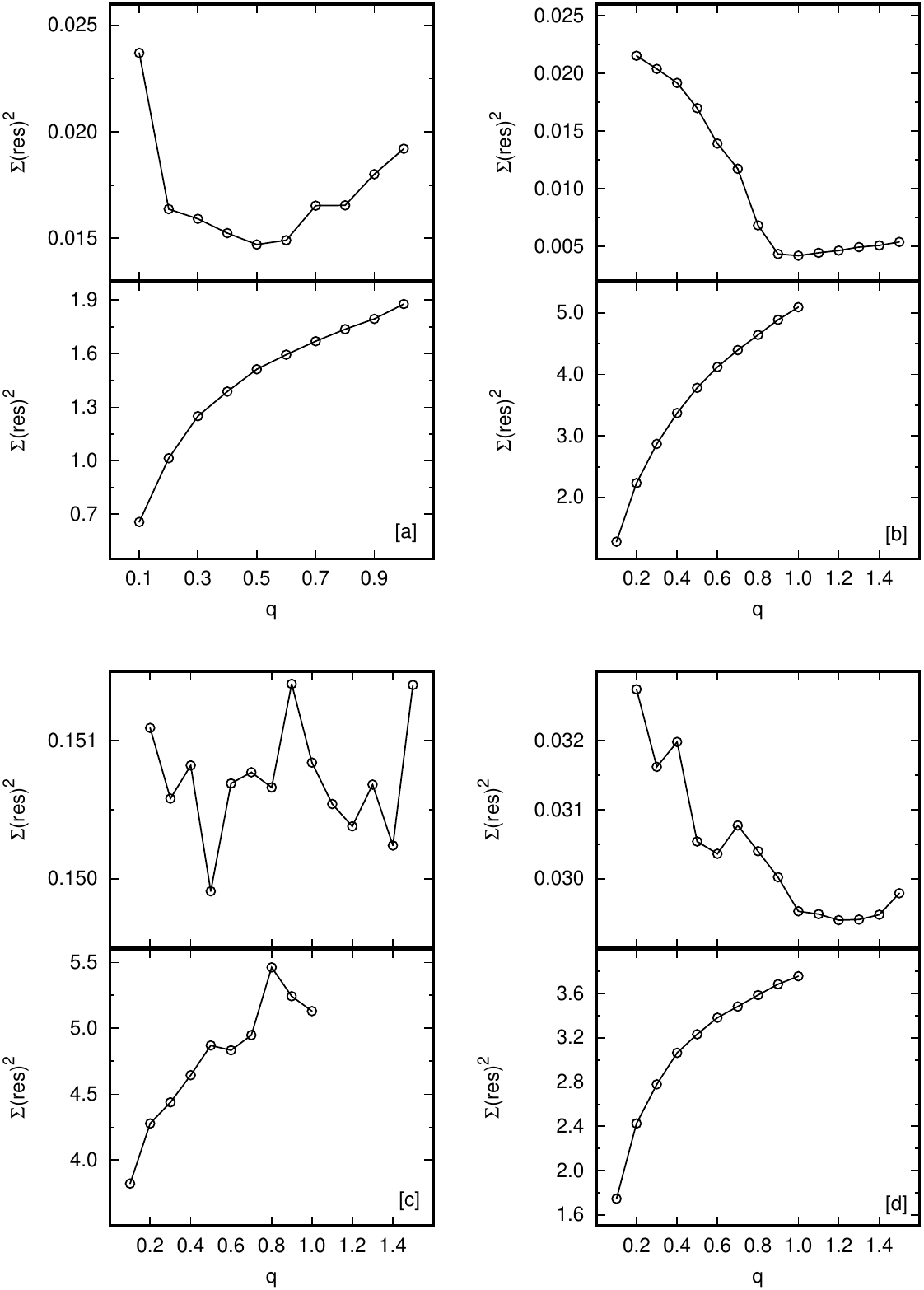}
   \caption{\label{figq}Results of the q-searches for CD-34~13220 (a), HD~295082 (b), TYC~ 6484-426-1 (c), and TYC~6527-2310-1 (d). The upper and lower panels refer to the solutions in detached and semi-detached assumptions.} 
   
   \end{figure}

The mass ratio is another significant parameter in the light curve analysis, just as temperature and orbital period mentioned in the previous paragraphs. The lack of any research related to the mass ratios of the systems in the literature pointed us to apply the $q$-search technique to the light curves in order to estimate mass ratio values. Although the shapes of the light curves and the orbital period values imply that the systems are detached binaries, we applied $q$-search to the systems' binned light curves covering 1500 data points in both detached and conventional semi-detached mode since there is no information about the morphological classes of the systems in the literature. The difference between $\Sigma(res)^2$ (sum of squared residuals) of the solutions with two assumptions confirmed that the systems are detached binaries, as it can easily be hypothesized from the light curve shapes even at the first glance. The results are given in Fig.~\ref{figq}.

We analysed the TESS light curves using the {\tt PHOEBE} (\citealt{pri05}) which uses the Wilson-Devinney method (\citealt{wil71}) to solve the light curve and obtain the stellar parameters from the input data. Following the results of the $q$-search, we applied analyses by assuming that the systems are detached binaries. The adjustable parameters are the inclination $i$, the temperature of the secondary component $T_2$, mass ratio $q$, the surface potential values of both components $\Omega_1$ and $\Omega_2$, and luminosity of the primary component $L_1$  during the solutions. The albedos ($A_{1}$= $A_{2}$=0.5) are calculated from \cite{ruc69} and the gravity darkening coefficients ($g_{1}$=$g_{2}$=0.32) for the systems was adopted from \cite{zei24} and \cite{luc67} by considering the granulation boundary for main sequence stars located at about F0 spectral type (\citealt{gra89}). Logarithmic limb darkening coefficients ($x_{1}$ and $x_{2}$) were adjusted from \cite{cla17} based on the initial temperatures of the components and given in Table~\ref{tablc}. In the table, the uncertainties in $T_1$ are listed as obtained in SED analyses, 2450000 is subtracted from the times of minimum lights and, moreover, TYC~ 6484 and TYC~6527 refer to TYC~ 6484-426-1 and TYC~6527-2310-1, respectively. The fill-out factors for the components, $f$, were also calculated by using the relation \scalebox{1.3}{$f=\frac{r}{r_{L}}$}, where \scalebox{1.3}{$r_{L}=\frac{0.49q^{\frac{2}{3}}}{0.6q^{\frac{2}{3}} + \ln(1+q^{\frac{1}{3}})}$}, the volume radius of the Roche lobe, given by \cite{egg83}.

\subsection{CD-34~13220}

\begin{figure}[htbp]
\centering
\includegraphics[scale=0.75]{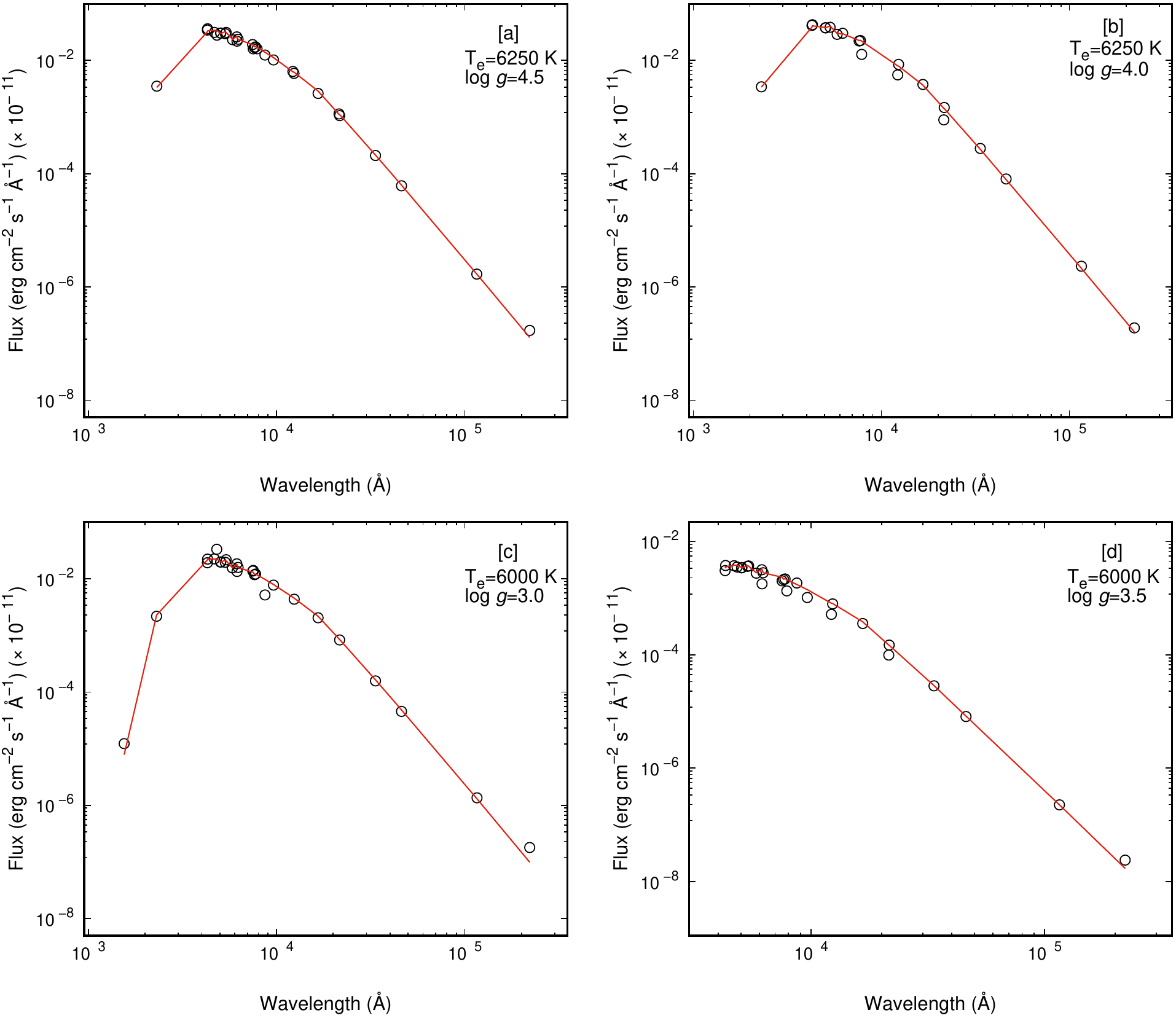}
\caption{Results of the SED analyses for CD-34~13220 (a), HD~295082 (b), TYC~ 6484-426-1 (c), and TYC~6527-2310-1 (d). Open circles indicate the photometric flux values from the VizieR database \citep{och00} and the solid (red in colored version) lines refer to the model fit. Final effective temperature and log~$g$ values are given on the upper right of each figure. Please note that the axes are on a logarithmic scale.}
\label{figsed}
\end{figure}

The temperature of the primary component was derived by SED analysis, as remarked in the previous section. The effective temperature interval was set 6000-6500~K during the SED analysis, comprising the temperatures, 6152~K of \cite{gai18} and 6345~K of \cite{sta19}. The log~$g$ interval was selected as 2.5-5.0 following TESS Input Catalog (TIC, \citealt{sta19}). The extinction, A$_{\nu}$=0.2246$^m$, was calculated using the Galactic Dust Reddening and Extinction interface of NASA/IPAC Infrared Science Archive\footnote{https://irsa.ipac.caltech.edu/applications/DUST/}. The results of the SED analysis were T$_e$=6250$\pm$125 and log~$g$=4.50$\pm$0.25. Therefore, the initial effective temperature of the primary was taken as 6250~K during the light curve solution. Fig.~\ref{figsed} shows the model fit of the SED analysis. 

\begin{figure}[htbp] 
   \centering
\includegraphics[scale=0.75]{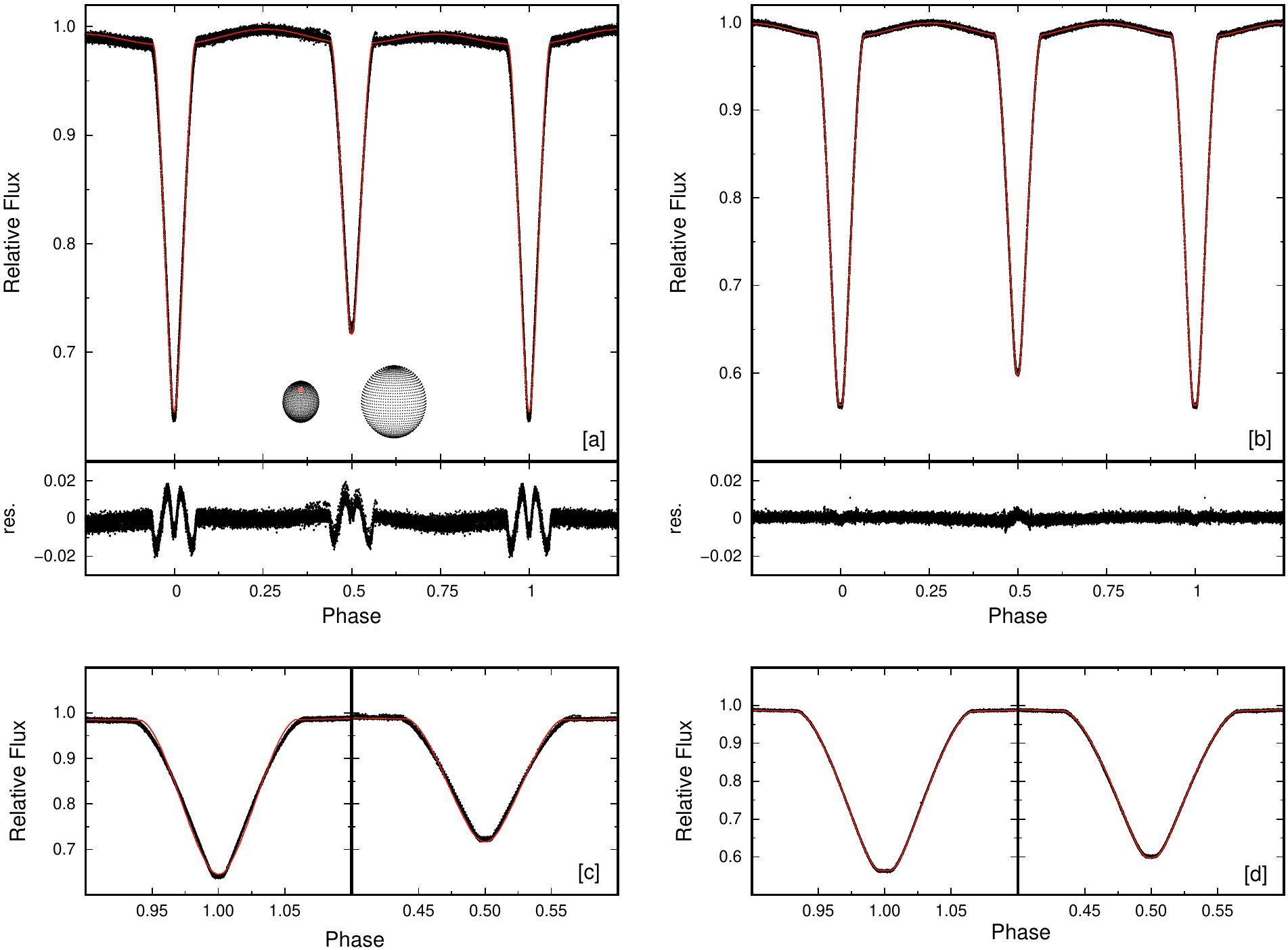}
\caption{Observational data with the synthetic light curves (red in colored version) of CD-34~13220 (a) and HD~295082 (b). The agreement of the fits are also shown at primary and secondary minimums for CD-34~13220 (c) and HD~295082 (d). The geometry of CD-34~13220 at phase 0.75 is also presented in (a) for the visibility of the spotted area (red) on the secondary component.} 
   \label{figlc1}
   \end{figure}

The light curve analysis was applied on 17766 data points in detached binary mode with the initial mass ratio value of 0.5 which was yielded by $q$-search. The orbital frequency was derived as 0.85551$d^{-1}$ by using {\tt PERIOD04} which corresponds to 1.1694~days. However, a more appropriate value, 1.167802~days, was found by calculating the average of differences between consecutive times of minimum lights (see Sec.~\ref{sec:lcan}). Moreover, as magnitude difference between two maxima of the light curve exists, the best solution was achieved adopting a spotted area on the secondary component with the parameters of co-latitude $\beta$=45~${{^\circ}}$, longitude $\lambda$=270~${{^\circ}}$, spot radius $r$=10~${{^\circ}}$, and the temperature factor $T_{f}$=0.9. The observations and the synthetic light curve are shown together in Fig.~\ref{figlc1}. The resulting light parameters are listed in Table~\ref{tablc}.

\begin{table}[htbp]
\caption{\label{tablc}Results of the light curve analyses. $r_1$ and $r_2$ refer to the fractional radii. The standard deviations, 3$\sigma$ for the last digits of light parameters are given in parentheses}
 \begin{tabular}{lcccc}
  \hline\noalign{\smallskip}
Parameter & CD-34~13220 & HD~295082 & TYC~6484 & TYC~6527\\
\hline                                                  
$i$ ${({^\circ})}$       & 82.38(2)   	& 87.78(1) &87.80(1) & 84.59(1)\\       
$q$                        & 0.443(1) & 1.001(1) &0.538(1)& 1.195(2)	\\         
$T_1$ (K)                  & 6250(125) & 6250(125) & 6000(125) & 6000(125)  	\\   
$T_2$ (K)                  & 5904(17) & 6055(10)&5970(24)& 5936(20)\\           
$\Omega _{1}$              & 3.473(9) & 6.622(2)&13.30(1) & 8.299(6)\\         
$\Omega _{2}$              & 3.642(6) & 5.459(3)&8.62(1)& 6.850(9)\\          
$\frac{L_1}{L_1 +L_2}$     & 0.588(1) & 0.4178(2)& 0.541(1)& 0.338(1)\\  
$r_1$                      & 0.199(3) & 0.1785(7)& 0.078(3) & 0.141(2) \\    
$r_2$                      & 0.189(5) &0.226(1)	& 0.073(3)& 0.202(2)\\ 
$f_1$                      & 0.64(1) &0.471(2) & 0.24(1)& 0.357(5) \\    
$f_2$                      & 0.61(2) &0.596(3) & 0.22(1)& 0.512(5)\\ 
$x_{1},~x_{2}$ 		   & 0.586, 0.601 & 0.583, 0.599 & 0.595, 0.611& 0.597, 0.597\\
T$_0$ (BJD) & 1657.169086(5)~ & 1470.39149(1)~ & 1441.23240(2)~ & 1468.81026(1)\\    
$P$~(days)   & 1.167802(3)  & 2.9430013(9) &  7.056833(1)  & 3.064855(2)\\
\hline
\end{tabular}

\end{table}

\subsection{HD~295082}
Our SED analysis resulted in temperature and surface gravity values of T$_e$=6250$\pm$125 and log~$g$=4.00$\pm$0.25. The initial interval of effective temperature was set 6000-7000~K during the analysis following \cite{gai18}, \cite{sta19} and F5 spectral type given by \cite{nes95}. The log~$g$ interval was 3.0-4.0 considering the value of 3.5 given by TIC (\citealt{sta19}) and the visual extinction was derived as 1.0618$^m$. The fit of the SED analysis plotted with the photometric data in Fig.~\ref{figsed}.

The light curve analysis was conducted with 14831 data points composing the light curve. During the analysis, the initial value of mass ratio was taken as 1.0, the result of our $q$-search. The orbital frequency was found to be 0.3352$d^{-1}$ (2.9832~days) and 2.9430013~days by using two methods in Sec.~\ref{sec:lcan} and set the latter one through the light curve solution process. The agreement between the theoretical light curve and observations is shown in Fig.~\ref{figlc1}. The final parameters found by the analysis are given in Table~\ref{tablc}.

\subsection{TYC~6484-426-1}
The SED analysis ended up with the values of T$_e$=6000$\pm$125 and log~$g$=3.00$\pm$0.25 based on the initial interval of 6000-6250~K and 3.0-4.0 for the effective temperature and surface gravity, respectively, following the values given by \cite{gai18} and \cite{sta19} (Fig.~\ref{figsed}).

\begin{figure}[htbp] 
   \centering
\includegraphics[scale=0.75]{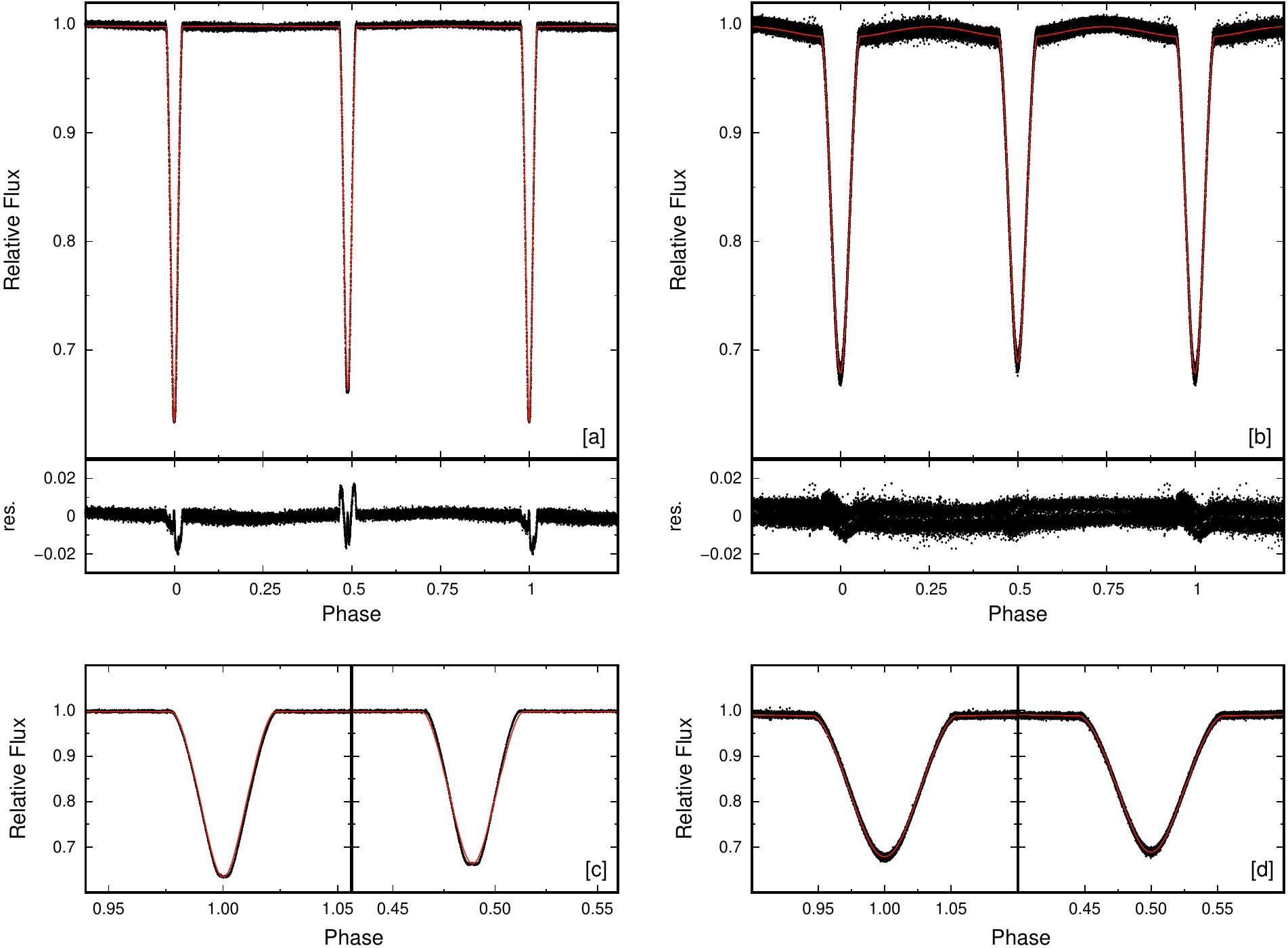}
   \caption{Same as Fig.~\ref{figlc1}, but for  TYC~ 6484-426-1 (a and c), and TYC~6527-2310-1 (b and d).}
   \label{figlc2}
   \end{figure}

The analysis was applied to the light curve which consists of 31924 data points. The initial mass ratio was 0.5 and the calculated value of the orbital period is 7.056833~days. Differently from our other three targets, the eccentricity was also set as a free parameter during the solution, and it was found to be $e$=0.0540(6). The synthetic light curve is plotted with the observational data in Fig.~\ref{figlc2} and the results are tabulated in Table~\ref{tablc}.

\subsection{TYC~6527-2310-1}
The temperature of the primary component was assumed to be 6000$\pm$125~K and the log~$g$ is 3.50$\pm$0.25, the results of the SED model fit. The SED analysis was applied by using the initial parameters of the effective temperature between 5750 and 6000~K and gravity between 3.0 and 5.0 by following values of \cite{gai18}, \cite{sch19}, \cite{sta19}. The result is plotted with photometric data in Fig.~\ref{figsed}.

31191 data points were analysed during the light curve solution with the initial mass ratio value of 1.2 from the $q$-search. The orbital period was derived as 3.064855~days. Results show that the secondary component is more massive and luminous than the primary as shown in Table~\ref{tablc}. Observational and synthetic light curves are plotted in Fig.~\ref{figlc2}.

We also investigated the solar-like oscillation candidacy of the star, as it was included in the asteroseismic target list of \cite{sch19}. Although the temperature of the target (6250~K) can be considered as close to solar temperature, the period analyses conducted with {\tt PERIOD04} did not result in any frequency close to 3mHz, the frequency of the highest amplitude oscillation of Sun. Therefore, we do not confirm the probable oscillational behavior of the star based on our analysis.

\section{Conclusion}
We present the first results proving that the systems in question are detached binaries. The effective temperature and log~$g$ values were estimated for primaries by using model fits from the SED analyses. The orbital periods were determined by applying frequency analysis and calculating the average of the differences between consecutive times of minimum lights. The $q$-search technique was used to estimate the initial values of mass ratios. The TESS light curve solutions indicated that the systems are detached eclipsing binaries. Unlike the other systems, a better fit was achieved by assuming a spotted area on the secondary component of CD-34~13220. We calculated the absolute parameters of the components of systems considering our analyses results by using {\tt{ABsParEB}} software (\citealt{lia15a}) and list in Table~\ref{tababs}. The masses of the primary components were estimated by using the stellar tracks for solar abundances of \cite{ber09} based on their log~$g$ and effective temperature. We conclude that the components of HD~295082 have very close mass and radius values while the secondary component of TYC~6527-2310-1 is more massive and luminous than the primary. The results implied that CD-34~13220 and TYC~ 6484-426-1 are typical detached binary systems.

\begin{table}[htbp]
\caption{\label{tababs}Absolute parameters of the systems. TYC~ 6484 and TYC~6527 refer to TYC~ 6484-426-1 and TYC~6527-2310-1, respectively. The standard errors are given in parentheses for the last digits}
 \begin{tabular}{lcccc}
  \hline\noalign{\smallskip}
Parameter & ~CD-34~13220~ & ~HD~295082~ & ~TYC~6484~ &~ TYC~6527\\
\hline
M$_1$ (M$_{\odot}$) &1.2& 1.3& 2.5& 1.8\\
M$_2$ (M$_{\odot}$) &0.532(1)& 1.301(1)& 1.345(3)& 2.151(4)\\
R$_1$ (R$_{\odot}$) &1.14(5)& 2.17(3)&1.9(5)& 2.0(1) \\
R$_2$ (R$_{\odot}$) &1.08(9)& 2.74(3)&1.8(6)& 2.90(8)\\
L$_1$ (L$_{\odot}$) &1.8(2)& 6.4(5)&4(2)& 4.8(7)\\
L$_2$ (L$_{\odot}$) &1.3(2)& 9.1(8)&4(2)& 9.4(9)\\
$a $ (R$_{\odot}$) & 5.736(4)& 12.165(5)&24.83(1)& 14.367(9) \\
\hline                                      
\end{tabular}
\end{table}

\begin{figure}[htbp] 
   \centering
\includegraphics[scale=0.8]{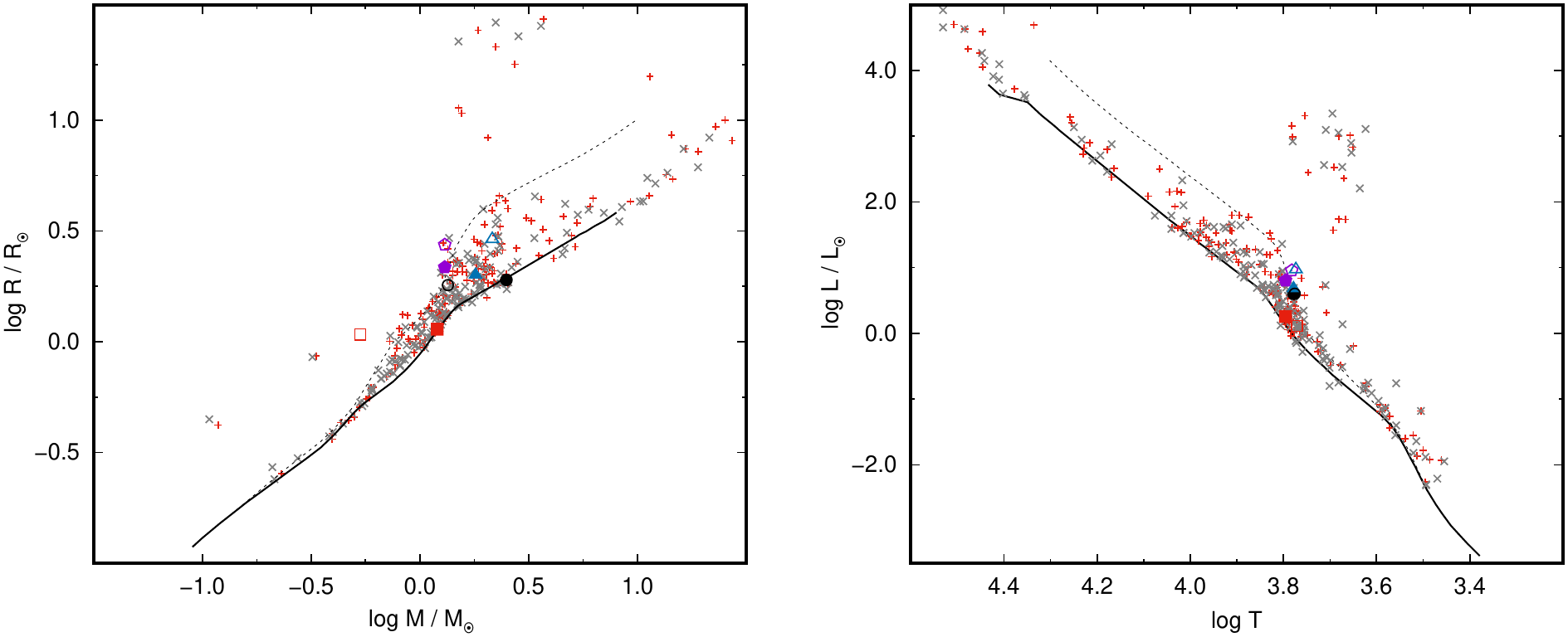}
\caption{Location of the components of the systems on the mass-radius plane (left) and the Hertzsprung-Russell diagram (right). Plus and crosses (pale colors; grey and tan in colored version) refer to the primary and secondary components of known detached systems given by \cite{sou15}. Filled signs remark the primary components of the systems, where the open signs denote the secondaries. Squares (red), pentagons (purple), circles (black), and triangles (blue) refer to the components of CD-34~13220, HD~295082, TYC~ 6484-426-1, and TYC~6527-2310-1, respectively. The data for ZAMS (thick solid line) and TAMS (dashed line) with Z=0.01 and Y=0.267 abundance are taken from \cite{bre12}.}
   \label{fighrmr}
   \end{figure}

\begin{figure}[htbp] 
   \centering
\includegraphics[scale=1.0]{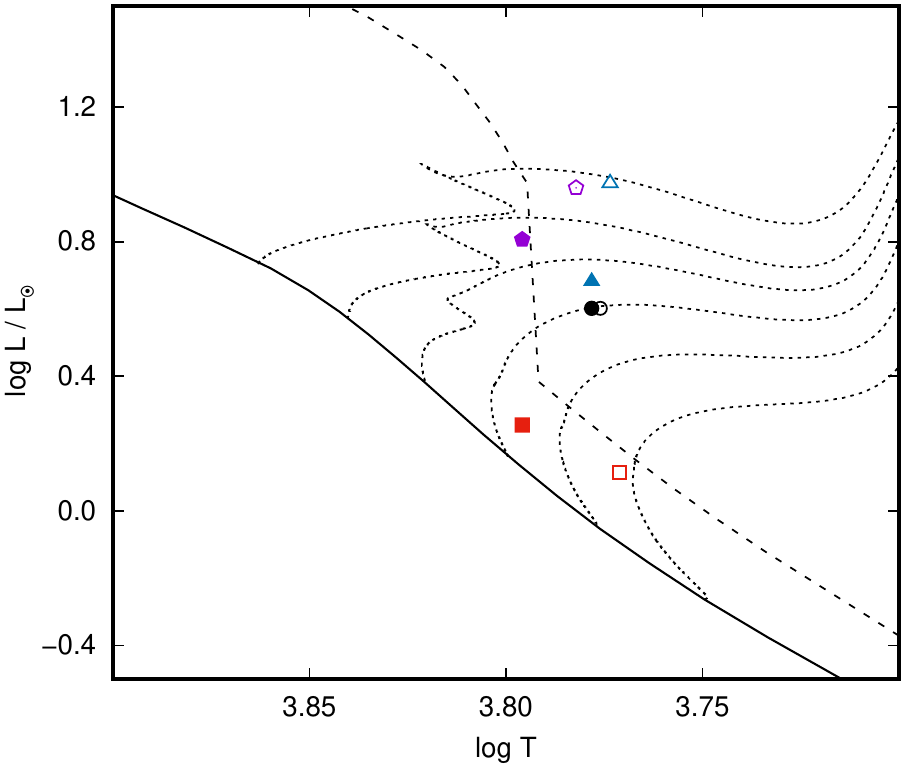}
\caption{Components of the systems on the Hertzsprung-Russell diagram with the evolutionary tracks for the stars with solar abundances and having the masses of 0.9~M$_{\odot}$, 1.0~M$_{\odot}$, 1.1~M$_{\odot}$,  1.2~M$_{\odot}$, 1.3~M$_{\odot}$ and 1.4~M$_{\odot}$ (dotted lines, from bottom to top). The other symbols are the same as those shown in Fig.~\ref{fighrmr}. The evolutionary tracks, ZAMS (solid line) and TAMS (dashed line) data for Z=0.01 and Y=0.267 composition are adopted from \cite{bre12}.}
   \label{figevo}
   \end{figure}

The four systems were compared to 162 well-known detached binaries cataloged by \cite{sou15}. In general, the locations of the components are in agreement with the other detached binary systems on the mass-radius plane and the Hertzsprung-Russell diagram (Fig.~\ref{fighrmr}). However, it must be pointed out that the secondary component of CD-34~13220 is found to be slightly less massive according to its radius. The components of TYC~6484-426-1 almost overlap on the Hertzsprung-Russell diagram because of their similarity in temperature and luminosity. The location of the systems were also plotted with the six evolutionary tracks of \cite{bre12} for masses from 0.9~M$_{\odot}$ to 1.4~M$_{\odot}$ in Fig.~\ref{figevo}. The components of the systems seem to follow the evolution of the stars, having the initial masses of between 0.9 and 1.4M$_{\odot}$. The figure also indicates that the secondary components of HD~295082 and TYC~6527-2310-1 are more evolved than the primaries which led us to consider the possibility of enhanced stellar winds discussed by \cite{tou88}. The authors focused on mass inversion in RS~CVn binaries and indicated that the secondary (originally less massive) component demonstrates tidally enhanced mass loss by the stellar wind. They also propose three different evolution scenarios based on Roche lobe overflow (RLOF) phase and period-mass ratio relation. Although the system Z Her used by authors in their model, shows some distinctive properties than our systems in terms of differences between radii and luminosities of the components, our derived mass ratios, and the mass values are agreed with the model star. According to their classification criteria based on final period and mass ratio, HD~295082 and TYC~6527-2310-1 are in the group of the systems whose RLOF stage begins at $q>0.7$ and causes rapid mass transfer through common envelope evolution. We remark that the enhanced stellar wind mechanism is within the realm of possibility in describing the evolved status of secondary components. However, the hypothesis needs to be confirmed by detailed stellar models which reveal the initial orbital period and the masses of the systems.

\begin{figure}[htbp] 
   \centering
\includegraphics[scale=0.85]{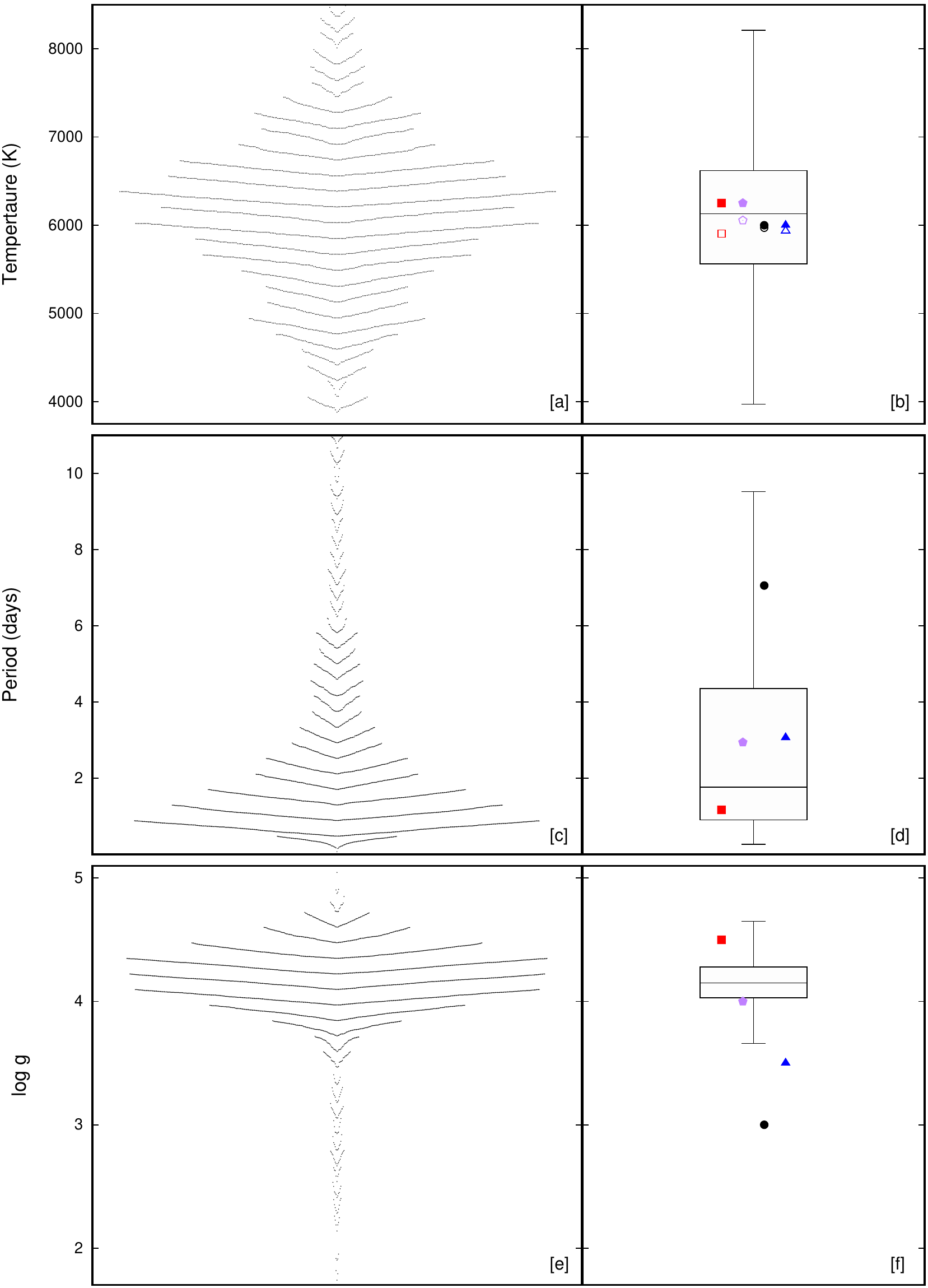}
\caption{Temperature (a), orbital period (c), and log~$g$ (e) distributions of EA-type LAMOST stars based on data from \cite{qia18} with their box plots (b, d, and e) where our targets are also flagged. Period values are plotted between 0 and 11 days (c) for the sake of clear visibility, although the box (d) calculated using the whole data interval. The symbols are the same as those shown in Fig.~\ref{fighrmr}. See text for details.}
   \label{figbox}
   \end{figure}

\begin{figure}[htbp] 
   \centering
\includegraphics[scale=0.85]{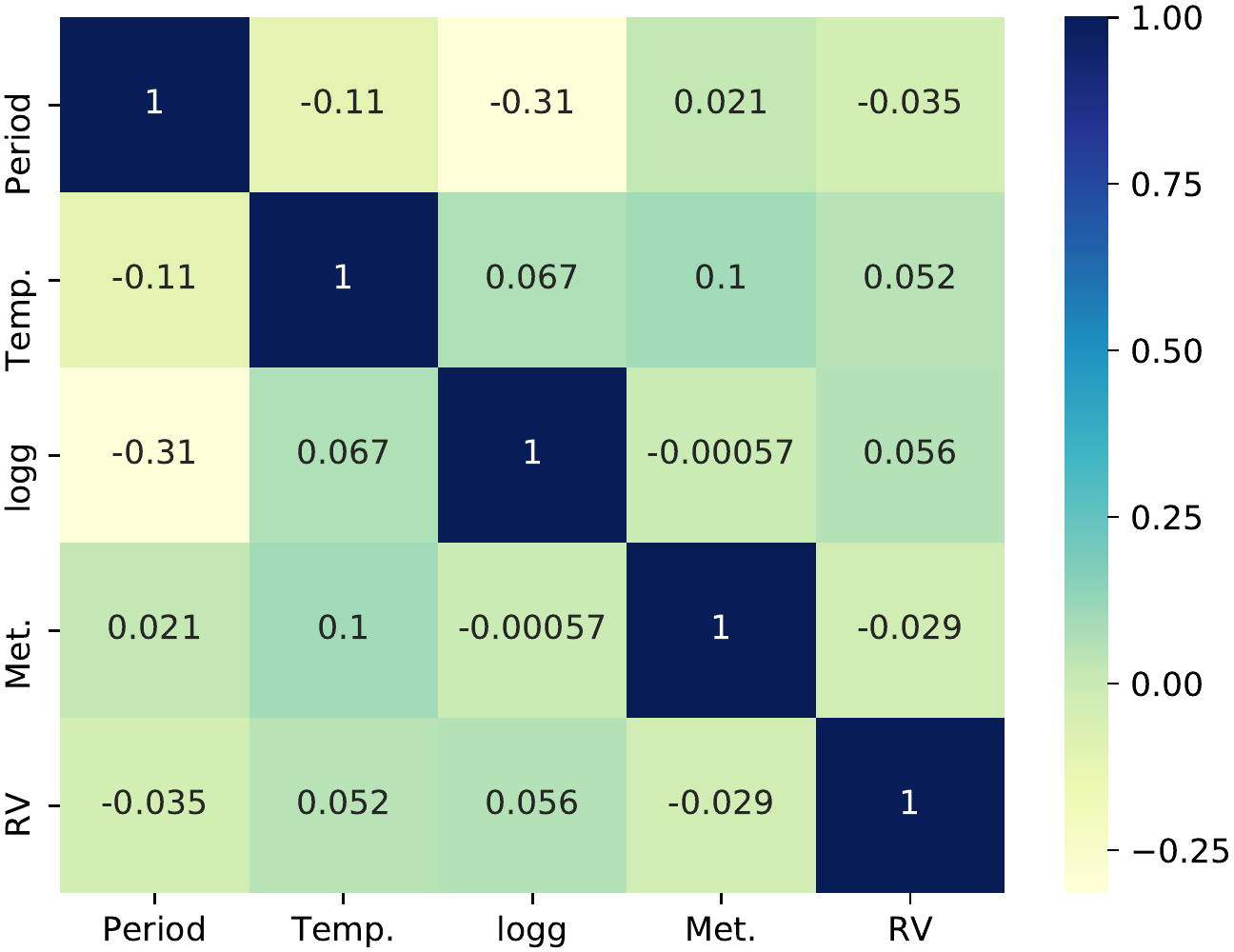}
\caption{Correlation matrix constructed based on parameter data of LAMOST EA stars given by \cite{qia18}. Temp., Met., and RV stand for temperature, metallicity, and radial velocity.}
   \label{figcor}
   \end{figure}

We compared our targets to LAMOST EA-type eclipsing binaries whose properties are given by \cite{qia18}. The distribution of the temperatures, orbital periods and log $g$ values of LAMOST EA binaries are given in Fig.~\ref{figbox}. We also drew box plots \citep{krz14} of those three distributions (right panels of Fig.~\ref{figbox}) to display the variation of the data and to examine its agreement with the parameters of our targets. The box plots indicate that the temperatures of the components of our systems are in good agreement with the same type of LAMOST stars. The orbital periods of CD-34~13220, HD~295082, and TYC~6527-2310-1 are in the interquartile range ($IQR$) as well. However, derived log~$g$ values deviate from the $IQR$s. The five-number summary of box plots is listed in Table~\ref{tabbox} with the number of data and $IQR$s which were calculated by using the equation $IQR = Q3-Q1$. Additionally, it is planned to inspect a relation among the given properties of LAMOST binaries (i.e. temperature, period, log~$g$, metallicity, and radial velocity) and compare to those of our targets. However, a correlation matrix (Fig.~\ref{figcor}) constructed using a Python code based on data of 2906 LAMOST EA binaries implied that the parameters are weakly connected (the correlation coefficients are between -0.31 and 0.1) to end up with reliable results. As a consequence, our results suggest that the systems are expected to follow two different evolutionary courses as remarked by \cite{qia18}; CD-34~13220 follow the evolutionary path from short-period EA system to less massive EW-type binary through AML. The other three long-period systems, on the other hand, are expected to evolve into long-period, massive EW-type binaries via case A mass transfer. 

\begin{table}[htbp]
\caption{\label{tabbox}Summary of the box plots of LAMOST EA-type binaries. Temperature and period values are given in K and days, respectively}

 \begin{tabular}{lccc}
  \hline\noalign{\smallskip}
 & ~Temperature~ & ~Period~~~& ~~log~$g$\\
\hline
Number of data &2956& 2906& 2956\\
Lower quartile ($Q1$) & 5560& 0.9063& 4.03\\
Median ($Q2$) &6130& 1.7671&4.15 \\
Upper quartile ($Q3$) &6620& 4.3571&4.28\\
Minimum& 3970&0.2909& 3.66\\
Maximum& 8210&9.5240& 4.65\\
$IQR $ & 1060&3.4508& 0.25 \\
\hline                                      
\end{tabular}
\end{table}

We plausibly conclude that the systems are eclipsing detached binary systems. Advanced spectroscopic observations are needed for a more precise determination of the binary properties of the systems and to confirm the results presented in this study.

\normalem
\begin{acknowledgements}
This paper includes data collected by the TESS mission. Funding for the TESS mission is provided by the NASA's Science Mission Directorate. Some/all of the data presented in this paper were obtained from the Mikulski Archive for Space Telescopes (MAST). STScI is operated by the Association of Universities for Research in Astronomy, Inc., under NASA contract NAS5-26555. Support for MAST for non-HST data is provided by the NASA Office of Space Science via grant NNX13AC07G and by other grants and contracts. This research has made use of the NASA/IPAC Infrared Science Archive, which is operated by the Jet Propulsion Laboratory, California Institute of Technology, under contract with the National Aeronautics and Space Administration. This publication makes use of VOSA, developed under the Spanish Virtual Observatory project supported by the Spanish MINECO through grant AyA2017-84089. VOSA has been partially updated by using funding from the European Union's Horizon 2020 Research and Innovation Programme, under Grant Agreement nº 776403 (EXOPLANETS-A). This research has made use of NASA's Astrophysics Data System. This research has made use of the VizieR catalogue access tool, CDS, Strasbourg, France.

\end{acknowledgements}
  
\bibliographystyle{raa}
\bibliography{msRAA-2021-0228.R1}

\begin{thebibliography}{44}
\providecommand\natexlab[1]{#1}
\providecommand\JournalTitle[1]{#1}

\bibitem[{Bayo} {et~al.}(2008)]{bay08}
{Bayo}, A., {Rodrigo}, C., {Barrado Y Navascu{\'e}s}, D., {et~al.} 2008, \aap,
  492, 277

\bibitem[{Bertelli} {et~al.}(2009)]{ber09}
{Bertelli}, G., {Nasi}, E., {Girardi}, L., \& {Marigo}, P. 2009, \aap, 508, 355

\bibitem[{Borucki} {et~al.}(2010)]{bor10}
{Borucki}, W.~J., {Koch}, D., {Basri}, G., {et~al.} 2010, Science, 327, 977

\bibitem[{Bressan} {et~al.}(2012)]{bre12}
{Bressan}, A., {Marigo}, P., {Girardi}, L., {et~al.} 2012, \mnras, 427, 127

\bibitem[{Cantat-Gaudin} {et~al.}(2018)]{can18}
{Cantat-Gaudin}, T., {Vallenari}, A., {Sordo}, R., {et~al.} 2018, \aap, 615,
  A49

\bibitem[{Claret}(2017)]{cla17}
{Claret}, A. 2017, \aap, 600, A30

\bibitem[{Cruzal{\`e}bes} {et~al.}(2019)]{cru19}
{Cruzal{\`e}bes}, P., {Petrov}, R.~G., {Robbe-Dubois}, S., {et~al.} 2019,
  \mnras, 490, 3158

\bibitem[{Cutri} {et~al.}(2003)]{cut03}
{Cutri}, R.~M., {Skrutskie}, M.~F., {van Dyk}, S., {et~al.} 2003, VizieR Online
  Data Catalog, II/246

\bibitem[{Eggleton}(1983)]{egg83}
{Eggleton}, P.~P. 1983, \apj, 268, 368

\bibitem[{Gaia Collaboration} {et~al.}(2018)]{gai18}
{Gaia Collaboration}, {Brown}, A.~G.~A., {Vallenari}, A., {et~al.} 2018, \aap,
  616, A1

\bibitem[{Gray} \& {Nagel}(1989)]{gra89}
{Gray}, D.~F., \& {Nagel}, T. 1989, \apj, 341, 421

\bibitem[{Guinan}(1993)]{gui93}
{Guinan}, E.~F. 1993, in Astronomical Society of the Pacific Conference Series,
  Vol.~38, New Frontiers in Binary Star Research, ed. K.-C. {Leung} \& I.-S.
  {Nha}, 1

\bibitem[{H{\o}g} {et~al.}(2000)]{hog00}
{H{\o}g}, E., {Fabricius}, C., {Makarov}, V.~V., {et~al.} 2000, \aap, 355, L27

\bibitem[{Kirk} {et~al.}(2016)]{kir16}
{Kirk}, B., {Conroy}, K., {Pr{\v{s}}a}, A., {et~al.} 2016, \aj, 151, 68

\bibitem[{Kovaleva} {et~al.}(2015)]{kov15}
{Kovaleva}, D., {Kaygorodov}, P., {Malkov}, O., {Debray}, B., \& {Oblak}, E.
  2015, Astronomy and Computing, 11, 119

\bibitem[Krzywinski \& Altman({2014})]{krz14}
Krzywinski, M., \& Altman, N. {2014}, {NATURE METHODS}, {11}, 119

\bibitem[{Kurucz}(1979)]{kur79}
{Kurucz}, R.~L. 1979, \apjs, 40, 1

\bibitem[{Kwee} \& {van Woerden}(1956)]{kwe56}
{Kwee}, K.~K., \& {van Woerden}, H. 1956, \bain, 12, 327

\bibitem[{Lenz} \& {Breger}(2005)]{len05}
{Lenz}, P., \& {Breger}, M. 2005, Communications in Asteroseismology, 146, 53

\bibitem[{Liakos}(2015)]{lia15a}
{Liakos}, A. 2015, in Astronomical Society of the Pacific Conference Series,
  Vol. 496, Living Together: Planets, Host Stars and Binaries, ed. S.~M.
  {Rucinski}, G.~{Torres}, \& M.~{Zejda}, 286

\bibitem[{Lucy}(1967)]{luc67}
{Lucy}, L.~B. 1967, \zap, 65, 89

\bibitem[{Morrison} {et~al.}(2001)]{mor01}
{Morrison}, J.~E., {R{\"o}ser}, S., {McLean}, B., {Bucciarelli}, B., \&
  {Lasker}, B. 2001, \aj, 121, 1752

\bibitem[{National Aeronautics and Space Administration}(1993)]{nat93}
{National Aeronautics and Space Administration}. 1993, Cape Photographic
  Durchmusterung, 3

\bibitem[{Nesterov} {et~al.}(1995)]{nes95}
{Nesterov}, V.~V., {Kuzmin}, A.~V., {Ashimbaeva}, N.~T., {et~al.} 1995, \aaps,
  110, 367

\bibitem[{Niemela}(2001)]{nie01}
{Niemela}, V. 2001, in Revista Mexicana de Astronomia y Astrofisica Conference
  Series, Vol.~11, Revista Mexicana de Astronomia y Astrofisica Conference
  Series, 23

\bibitem[{Ochsenbein} {et~al.}(2000)]{och00}
{Ochsenbein}, F., {Bauer}, P., \& {Marcout}, J. 2000, \aaps, 143, 23

\bibitem[{Pojmanski}(2002)]{poj02}
{Pojmanski}, G. 2002, \actaa, 52, 397

\bibitem[{Pr{\v{s}}a} \& {Zwitter}(2005)]{pri05}
{Pr{\v{s}}a}, A., \& {Zwitter}, T. 2005, \apj, 628, 426

\bibitem[{Qian} {et~al.}(2018)]{qia18}
{Qian}, S.~B., {Zhang}, J., {He}, J.~J., {et~al.} 2018, \apjs, 235, 5

\bibitem[{Qian} {et~al.}(2020)]{qia20}
{Qian}, S.-B., {Zhu}, L.-Y., {Liu}, L., {et~al.} 2020, Research in Astronomy
  and Astrophysics, 20, 163

\bibitem[{Qian} {et~al.}(2019)]{qia19}
{Qian}, S.-B., {Shi}, X.-D., {Zhu}, L.-Y., {et~al.} 2019, Research in Astronomy
  and Astrophysics, 19, 064

\bibitem[{Ricker} {et~al.}(2015)]{ric15}
{Ricker}, G.~R., {Winn}, J.~N., {Vanderspek}, R., {et~al.} 2015, Journal of
  Astronomical Telescopes, Instruments, and Systems, 1, 014003

\bibitem[{R{\"o}ser} {et~al.}(1994)]{ros94}
{R{\"o}ser}, S., {Bastian}, U., \& {Kuzmin}, A. 1994, \aaps, 105, 301

\bibitem[{Ruci{\'n}ski}(1969)]{ruc69}
{Ruci{\'n}ski}, S.~M. 1969, \actaa, 19, 245

\bibitem[{Ruiz-Dern} {et~al.}(2018)]{rui18}
{Ruiz-Dern}, L., {Babusiaux}, C., {Arenou}, F., {Turon}, C., \& {Lallement}, R.
  2018, \aap, 609, A116

\bibitem[{Schofield} {et~al.}(2019)]{sch19}
{Schofield}, M., {Chaplin}, W.~J., {Huber}, D., {et~al.} 2019, \apjs, 241, 12

\bibitem[{Southworth}(2015)]{sou15}
{Southworth}, J. 2015, in Astronomical Society of the Pacific Conference
  Series, Vol. 496, Living Together: Planets, Host Stars and Binaries, ed.
  S.~M. {Rucinski}, G.~{Torres}, \& M.~{Zejda}, 164

\bibitem[{Stassun} {et~al.}(2019)]{sta19}
{Stassun}, K.~G., {Oelkers}, R.~J., {Paegert}, M., {et~al.} 2019, \aj, 158, 138

\bibitem[{Tanenbaum} \& {Jenkins}(2018)]{tan18}
{Tanenbaum}, P., \& {Jenkins}, J.~M. 2018, TESS Science Data Products
  Description Document, EXP-TESS-ARC-ICD-0014 Rev D

\bibitem[{Thome}(1994)]{tho94}
{Thome}, M.~J. 1994, VizieR Online Data Catalog, I/114

\bibitem[{Tout} \& {Eggleton}(1988)]{tou88}
{Tout}, C.~A., \& {Eggleton}, P.~P. 1988, \mnras, 231, 823

\bibitem[{von Zeipel}(1924)]{zei24}
{von Zeipel}, H. 1924, \mnras, 84, 665

\bibitem[{Wilson} \& {Devinney}(1971)]{wil71}
{Wilson}, R.~E., \& {Devinney}, E.~J. 1971, \apj, 166, 605

\bibitem[{Zhao} {et~al.}(2012)]{zha12}
{Zhao}, G., {Zhao}, Y.-H., {Chu}, Y.-Q., {Jing}, Y.-P., \& {Deng}, L.-C. 2012,
  Research in Astronomy and Astrophysics, 12, 723

\end{thebibliography}

\end{document}